\documentclass[titlepage,12pt]{article}
\usepackage{amsmath,amsfonts,graphicx}
\title{Counting {Hamiltonian} cycles\\
 on planar random  lattices}
\author{
{Saburo Higuchi}\thanks{e-mail: hig@rice.c.u-tokyo.ac.jp}\\[0.5ex]
{\it Department of Pure and Applied Sciences,} \\
{\it The University of Tokyo} \\
{\it  Komaba, Meguro, Tokyo 153-8902, Japan}
}
\date{January 29, 1998 \ \   
      UT-KOMABA/98-3,\ \ 
      cond-mat/9801307}

\begin{document}
\begin{titlepage}
\thispagestyle{empty}
  \begin{flushright}
    January 1998\\
    UT-KOMABA/98-3\\
    cond-mat/9801307
  \end{flushright}
\vspace*{\fill}

  \begin{center}
{\Huge\bf Counting {Hamiltonian} cycles\\[1ex]
 on planar random  lattices}
\vspace*{\fill}

{\Large Saburo Higuchi}%
\footnote{e-mail: \texttt{hig@rice.c.u-tokyo.ac.jp}}\\[0.5ex]

{\it Department of Pure and Applied Sciences,} \\
{\it The University of Tokyo} \\
{\it  Komaba, Meguro, Tokyo 153-8902, Japan}
  \end{center}
\vspace*{\fill}

\vspace*{\fill}
\begin{center}
{\large\bf Abstract}
\end{center}
\smallskip

A Hamiltonian cycle of a graph is a closed path which visits each of
the vertices once and only once.  In this article, Hamiltonian cycles
on planar random lattices are considered. The generating function for
the number of Hamiltonian cycles is obtained and its singularity is
studied.  Relation to two-dimensional quantum gravity is discussed.
\vspace*{5ex}
\end{titlepage}

Hamiltonian cycles have often been used to
model collapsed polymer globules\cite{ClJa:polymerE}. 
A Hamiltonian cycle of a graph is a closed path
which visits each of the vertices once and only once.
The number of Hamiltonian cycles on a graph corresponds to the
entropy of polymer 
system on it in collapsed but disordered phase. 
There  can be no polynomial time algorithms
to determine whether the number is zero or not 
which work for arbitrary graphs\cite{GaJo:NP-completeness}.
Even for regular graphs (lattices), 
the number of Hamiltonian cycles is not obtained exactly except for
a few well-behaved cases\cite{Kasteleyn,Lieb,Suzuki:regular,BaSuYu:honyeycomb} 

In the present work, I consider the problem of 
counting the number of Hamiltonian cycles on planar random lattices, or
planar random fat graphs. I obtain the exact generating function for
the number.

Let $S^n$ be the set of all planar trivalent fat graphs 
with $n$ vertices  possibly with multiple edges and self-loops.
Graphs that are isomorphic are identified.
The set $\tilde{S}^n$ is the labeled version of $S^n$,
namely, vertices of $\tilde{G}\in\tilde{S}^n$ are labeled as
$1,\ldots,n$ 
and $\tilde{G}_1, \tilde{G}_2\in\tilde{S}^n$ are considered identical
only if a graph isomorphism preserves labels.
The symmetric group of degree $n$
naturally acts on $\tilde{S}^n$ by label permutation.
The stabilizer subgroup of $\tilde{G}$ is called
the automorphism group $\mathrm{Aut\ }G$.

\begin{figure}[tb]
  \begin{minipage}{0.47 \textwidth}
  \begin{center}
    \leavevmode
\includegraphics[scale=0.4]{fatgraph.eps}        
\caption{An example of planar trivalent fat graph $G$ with $n=4$.
  A Hamiltonian cycle is drawn in thick line.
  For this graph, $\mathcal{H}(G)=4!$ because I distinguish the
  directions and the base points of the cycle.
\label{fatgraph}}
  \end{center}
  \end{minipage}
\hspace{\fill}
  \begin{minipage}{0.47 \textwidth}
  \begin{center}
    \leavevmode
\includegraphics[scale=0.4]{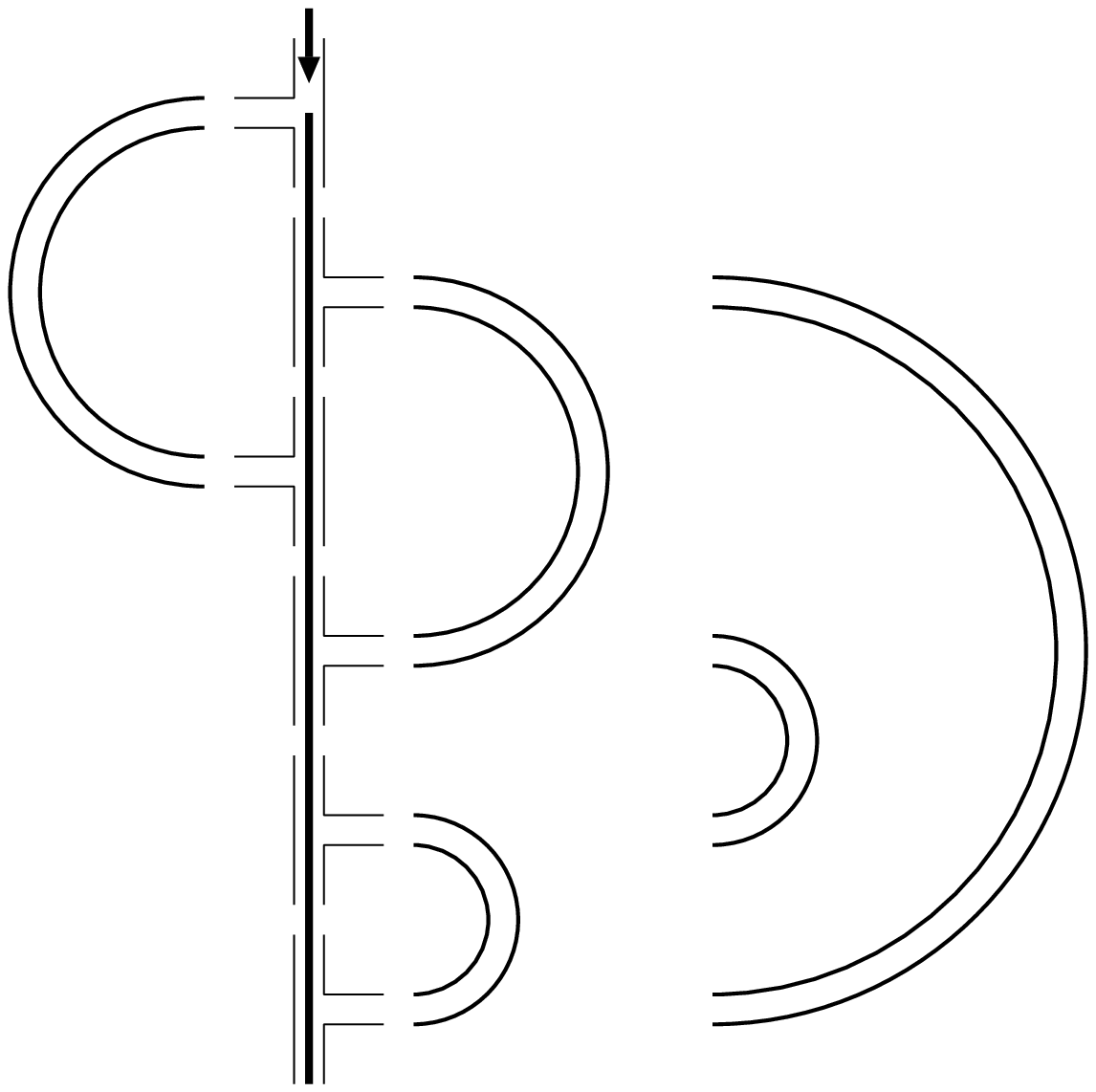}        
\caption{Two of contributing diagrams for $n=6$.
  The sequence of T's in the center is obtained by walking along 
  the Hamiltonian cycle (the thick line in the T's).
  Openings in both sides should be connected by arcs.
  On the left hand side, 
  there is a single possibility because $A_2=1$. On the right hand side,
  $A_4=2$ and the two possible patterns are drawn.
  \label{disks}}
  \end{center}
\end{minipage}
\end{figure}

A Hamiltonian cycle of a labeled graph $\tilde{G}\in \tilde{S}^n$ is a
directed closed path (consecutive distinct edges connected at vertices) 
which visits each of $n$ vertices exactly once.
 Hamiltonian cycles are understood as furnished with a direction
and a base point.
The number of Hamiltonian cycles of $\tilde{G}$ is denoted by 
$\mathcal{H}(G)$ because it is independent of the way of labeling.
See figure \ref{fatgraph} for an example.

The quantity I study in this work is 
\begin{equation}
 F_n = \sum_{G\in S^n} \mathcal{H}(G)\frac{1}{\#\mathrm{Aut\ } G}
 \label{f_n}
\end{equation}
and the generating function
\begin{equation}
  F(p) = \sum_{n=0}^\infty p^n  F_n.
\end{equation}

It is illuminating and is useful to rewrite $F_n$ as 
the number of isomorphism classes of the pair (graph, Hamiltonian cycle):
\begin{equation}
  F_n = \# ( \{ (\tilde{G}, C^n) | 
\tilde{G}\in \tilde{S}^n, C^n: \text{Hamiltonian cycle on }\tilde{G} \}/\sim),
  \label{num_of_pairs}
\end{equation}
where $(\tilde{G}_1,C^n_1)\sim(\tilde{G}_2,C^n_2)$ 
if and only if
$G_1$ and $G_2$ are isomorphic (forgetting the labels)
and the isomorphism maps $C^n_1$ onto $C^n_2$ 
with the direction and the initial point preserved.
Eq. \eqref{num_of_pairs} implies that $F_n$ is an integer though the
definition  \eqref{f_n} involves a fraction.

The equality \eqref{num_of_pairs} can be shown as follows.
\begin{align}
&  \# \{ (\tilde{G}, C^n) | 
   \tilde{G}\in \tilde{S}^n, C^n: \text{Hamiltonian cycle on }\tilde{G}\}/\sim
\notag\\
=& \frac{1}{n!} \sum_{\tilde{G}\in\tilde{S}^n} \# 
\{C^n | C^n: \text{Hamiltonian cycle on }\tilde{G}\}
\notag\\
=& \frac{1}{n!} \sum_{\tilde{G}\in\tilde{S}^n} \mathcal{H}(G)\notag\\
=& \frac{1}{n!} \sum_{G\in S^n} \frac{n!}{\#\mathrm{Aut\ }G}\mathcal{H}(G).
\end{align}
The first equality follows from the fact that there are $n!$
inequivalent labeling of $(G,C)$.
In the last line, the definition of $\mathrm{Aut\ }G$ is made use of.

To compute $F(p)$,
the method used in
refs. \cite{DuKo:random,DuKo:spectra,Ishibashi:2dstring} 
is followed.
One walks along the Hamiltonian cycle in the specified direction starting
from the base point and records the order of right and left turns.
Then one obtains a diagram consisting of T's 
as depicted in the center of figure \ref{disks}.
A cycle corresponds to exactly 2 out of $2^n$ diagrams consisting of
$n$ T's. The two are mirror images of each other.
There are $\binom{n}{k}$ diagrams that have $k$ openings on the left
hand side and $n-k$ on the right hand side.
To reproduce the graph completely, one has to connect the $n$ openings
pairwisely. 
There should be no connection between the right hand side and the left 
hand side because the cycle divides the sphere into two disks.
The connection pattern should be able to be drawn on a disk faithfully,
\textsl{i.e.} without intersection.
The number of ways of contracting $k$ objects on a disk is denoted by $A_k$.
Then one has
  \begin{equation}
  F_n  =  \sum_{k=0}^n \frac12 \binom{n}{k}     A_k     A_{n-k}.
                             \label{binom}
\end{equation}
The factor $1/2$ comes form the identification of the mirror images.
There is no double count besides it.
The number $A_k$ has been 
obtained by Br\'ezin \textsl{et. al.} using the gaussian matrix model
\cite{BrItPaZu:planar}:
  \begin{equation}
 A_k=\frac1{2\pi}\int^2_{-2}dx \sqrt{4-x^2}\quad x^{k}
   =
 \begin{cases}
   0 & (k:\text{odd}),\\
   \frac{k!}{(\frac{k}{2})!(\frac{k}{2}+1)!}&  (k:\text{even}).
 \end{cases}
  \end{equation}
Plugging this into eq.\eqref{binom}, one has
\begin{align}
F(p)       
      =& \sum_{n=0}^\infty  p^n\frac12
\frac{1}{2\pi} \int^2_{-2} dx \sqrt{4-x^2}
\frac{1}{2\pi} \int^2_{-2} dy \sqrt{4-y^2}
(x+y)^n\notag\\
      =&  \frac12
\frac{1}{2\pi} \int^2_{-2} dx \sqrt{4-x^2}
\frac{1}{2\pi} \int^2_{-2} dy \sqrt{4-y^2}
\frac{1}{1-A(x,y;p)},
\label{integral}
  \end{align}
where
\begin{equation}
  A(x,y;p)=px+py \label{A}
\end{equation}
By using 
\begin{equation}
\frac{1}{1-px-py} = \int^\infty_0  d\alpha \quad e^{(-1+p(x+y))\alpha},
\end{equation}
one  obtains
\begin{equation}
  F(p)=\frac{1}{2 p^2} \int^\infty_0 d\alpha \quad \frac{e^{-\alpha}}{\alpha^2}
    (I_1(2p\alpha))^2,
\end{equation}
where $I_1$ is the modified Bessel function of the first order.

From the representation \eqref{integral}, it is apparent that $F(p)$
diverges at $p=1/4$. The singularity of $F(p)$ brings 
information of large-$n$ behavior of $F_n$.
By studying it one can inspect the properties of  large graphs.
One can show  that $F(p)$  has a singularity
\begin{equation}
F(p)
=\frac{1}{16\pi p^{3}} (4p-1)^2 \log|4p-1|+\text{regular terms}.
\label{singular}
\end{equation}
by examining  \eqref{integral}.
This result is in accord  with the scaling behavior of a polymer loop  with a 
base point on planar random surfaces in the dense phase 
\begin{equation}
 F(p) \sim (p-\tfrac14)^2 
\end{equation}
obtained in  \cite{DuKo:random}.

Each planar trivalent fat graph corresponds to a 
triangulation of a sphere under the dual transformation.
Thus $F(p)$ is the partition function of two-dimensional quantum gravity
with an unusual weight.
Triangulations that do not admit a Hamiltonian cycle are excluded
from the path integral. On the other hand, ones admitting Hamiltonian
cycles are weighted by $\mathcal{H}(G)$.
Eq. \eqref{singular} shows that the string susceptibility exponent
$\gamma_{\text{string}}$ for this system is zero. 
Because there is a base point on the Hamiltonian cycle, one has a
local degree of freedom which gives rise to a factor proportional to
the area of the surface. 
In this viewpoint, 
one should compare the value $\gamma_{\text{string}}=0$ with scaling
exponent of  $c=-2$ quantum gravity with a puncture operator inserted.

This system can be generalized by introducing two parameters $p_1$
and $p_2$ as
\begin{equation}
 F(p_1,p_2)= \sum_{k,m=0}^\infty p_1^k p_2^m \sum_{G\in S^{k+m}}
\frac{1}{\#\mathrm{Aut\ } G} \mathcal{H}_{k,m}(G),
\label{turns}
\end{equation}
where
$\mathcal{H}_{k,m}(G)$ denotes 
the number of Hamiltonian cycle with $k$ right turns and $m$ left turns.
Then straightforward calculation shows
\begin{align}
F(p_1,p_2)       
      =&  \frac12
\frac{1}{2\pi} \int^2_{-2} dx \sqrt{4-x^2}
\frac{1}{2\pi} \int^2_{-2} dy \sqrt{4-y^2}
\frac{1}{1-p_1x-p_2y}\notag\\
=&\frac{1}{2 p_1p_2} \int^\infty_0 d\alpha \frac{e^{-\alpha}}{\alpha^2}
    I_1(2p_1\alpha)I_1(2p_2\alpha)\notag\\
=&\frac{1}{16\pi p_1^{3/2} p_2^{3/2}}
(2p_1+2p_2-1)^2 \log|2p_1+2p_2-1|+\text{regular terms}.
\end{align}
The function $F(p_1,p_2)$ is not only symmetric but 
with a  singularity  described in terms of $p_1+p_2$ only.
This implies that the sum is dominated by the cycles with an equal number
of right and left turns. 
The fluctuating geometry does not change the dominance of the maximum
of the factor $\binom{k+m}{k}$ at $k=m$.
I comment that in \eqref{turns}
the integer $\frac{2}{3}\pi(k-m)$ can be considered as the  holonomy
\cite{Ishibashi:2dstring}
of the cycle if one regards the system as a quantum gravity.

This solution can be extended to  planar $q$-valent fat graph.
One  should just replace $A(x,y;p)$ in \eqref{integral}
by $A(x,y;p)= p \sum_{s=0}^{q-2} x^s y^{q-2-s}$.
One can even consider the system with 
\begin{equation}
  A(x,y;\{p_{i,j}\})=\sum_{i,j=0}^N  p_{i,j} x^i y^j.
  \label{general}
\end{equation}
The system now describes Hamiltonian cycles on graphs 
consisting of $(i+j+2)$-valent vertices with various $i,j$.
The weight  $p_{i,j}$ corresponds to the
$2\pi\times\frac{i+1}{i+j+2}$-turn on a $(i+j+2)$-valent vertex. 
For example, if one takes  
$A(x,y;p_1,p_2)=p_1x^2 + p_2xy + p_1y^2$, 
Hamiltonian cycles with bending energy parameterized
by $p_1/p_2$ are realized on a random 4-valent lattice.

When one generalizes the system as \eqref{general},
the critical line of  $F(\{p_{i,j}\})$ is
\begin{equation}
A(x,y,\{p_{i,j}\})=1. \label{critical_line}
\end{equation}
Assume that there exists $q_0-2\in \mathbb{N}$ such that
\begin{gather}
  p_{i,j}=0 \text{\ if } i+j>q_0-2,\\
  p_{i,j}\neq0 \text{\ if\ } i+j=q_0-2.
\end{gather}
This amounts to assuming that the largest coordination number of the
graph is $q_0$ and the  cycle can turn arbitrarily on the vertex.
From direct evaluation of singular part of \eqref{integral} with
\eqref{general}, one can show that the singularity of $F(\{p_{i,j}\})$ is still
of the form 
\begin{equation}
F(\{p_{i,j}\})\sim(\delta p)^2 \log \delta p,
\end{equation}
where $\delta p$ is a function of $p_{i,j}$'s which vanishes linearly
on the critical curve. 

\begin{figure}[tbp]
  \begin{center}
    \leavevmode
\includegraphics[scale=0.6]{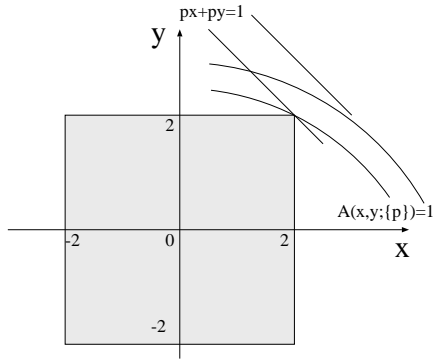}        
\caption{
The shaded region is the domain of integration \eqref{integral}.
The singularity occurs when the line $px+py=1$  touches this shaded region.
For the generalized case \eqref{general}, the line is replaced with an 
algebraic curve $1=A(x,y,\{p_{i,j}\})$ but the local
situation around $(x,y)=(2,2)$ is the same as the original case.
\label{xyplane}
}
  \end{center}
\end{figure}

This fact can be understood as follows.
Let us look at  the original case \eqref{integral} with \eqref{A}.
The range of integration of $x,y$ is  the square domain $-2\le x,y \le2$. 
The singularity occurs when the line $1-px-py=0$  touches this domain
as depicted in Fig. \ref{xyplane}.
It occurs on $(x,y)=(2,2)$ assuming $p>0$.
In the  case \eqref{general}, the critical curve is now 
an algebraic curve \eqref{critical_line}.
It intersects  the square at $(x,y)=(2,2)$ transversally.
The curve can be approximated by a straight line
with a slope when one is interested in critical behaviors.
The slope is the only relevant parameter at the  criticality.

In conclusion, I have obtained the generating function for the number
of Hamiltonian cycles on planar random  lattices and have considered
the limit of large graphs. 
\medskip

I thank Mitsuhiro Kato for useful discussions.
This work is supported by CREST from Japanese Science and Technology
Corporation. 
\medskip

While this paper is being typed, a preprint by 
B. Eynard, E. Guitter, and C. Kristjansen 
has appeared \cite{EyGuKr:hamiltonian}.
It contains interesting results closely related to the present paper.
Their definition of Hamiltonian cycle is slightly different from that
taken here; 
they do not associate  directions and  base points to cycles.

\bibliographystyle{physlett}

\bibliography{shrtjour,mrabbrev,polymer,2dgrav}

\end{document}